\documentclass[pre,aps,superscriptaddress,showpacs,twocolumn]{revtex4}
\usepackage{epsfig}
\usepackage{bm}

\begin{document}

\def\B.#1{{\bm #1}}
\def\C.#1{{\cal #1}}
\title{Active and Passive Fields in Turbulent Transport: the Role of
Statistically Preserved
Structures}
\author{Emily S.C. Ching}
\affiliation{Department of Physics, The Chinese University of Hong Kong,
Sha Tin, Hong Kong}
\author{Yoram Cohen}
\author{Thomas Gilbert}
\author{Itamar Procaccia}
\affiliation{Dept. of Chemical Physics, The Weizmann Institute
of Science, Rehovot 76100, Israel}
\date{\today}
\begin{abstract}
We have recently proposed that the statistics of {\em active}
fields (which affect the velocity field itself) in
well-developed turbulence
are also dominated by the Statistically Preserved Structures
of auxiliary {\em passive} fields which are
advected by the same velocity field.
The Statistically Preserved Structures are eigenmodes
of eigenvalue 1 of an appropriate propagator of the
decaying (unforced) passive field, or equivalently,
the zero modes of a related operator.
In this paper we investigate
further this surprising finding via two examples, one akin to turbulent
convection in which the temperature is the active scalar, and the other
akin to magneto-hydrodynamics in which the magnetic field is the active
vector. In the first example, all the even correlation functions
of the active and passive fields exhibit identical
scaling behavior.  The second example appears at first sight to be a
counter-example: the statistical objects of the active and passive fields
have entirely different scaling exponents. We demonstrate
nevertheless that the Statistically Preserved Structures
of the passive vector dominate again the statistics of the active
field, except that due to a dynamical conservation law the amplitude
of the leading zero mode cancels exactly. The active vector is then
dominated by the sub-leading zero mode of the passive vector. Our work
thus suggests that the statistical properties of active fields in
turbulence can be understood with the same generality as those of passive
fields.
\end{abstract}
\maketitle
\section{Introduction}
\label{intro}

The aim of this paper is to address the statistical physics of so called
``active" fields in developed fluid turbulence. These are fields that
differ
from the
fundamental fluid velocity field $\B.u(\B.r,t)$, but that interact with
the
velocity field
in an essential way, for example effecting a significant change in the
scaling exponents
of the velocity correlation functions from the classical Kolmogorov
exponents.
For the sake of concreteness we will focus on two generic examples with
very different interactions between the active and the velocity fields.

The first is thermal turbulent convection, in which the temperature field
$T(\B.r,t)$ is driving the velocity field through buoyancy effects. In the
Boussinesq
approximation the temperature equation reads like a standard forced
scalar advection problem,
\begin{equation}
\frac{\partial T(\B.r,t)}{\partial t}\!+\!\B.u(\B.r,t)\cdot
\B.\nabla T(\B.r,t)\!=\!\kappa \nabla^2 T(\B.r,t)\!+\!f(\B.r,t)  .
\label{active}
\end{equation}
Here $\kappa$ is the thermal diffusivity
and $f(\B.r,t)$ is a white random force of zero mean
with compact support in $\B.k$-space, acting on the largest
scales of the order of the outer scale $L$ only.
The velocity field is affected by the temperature. For
an incompressible fluid of unit density \cite{71MY} (dropping
the dependence on $(\B.r,t)$ for brevity),
\begin{equation}
\frac{\partial \B.u}{\partial t}+\B.u\cdot
\B.\nabla \B.u=-\B.\nabla p+\nu \nabla^2 \B.u
+\alpha g T\hat z \ .
\label{Boussinesq}
\end{equation}
Here $p$, $\nu$, $\alpha$, $g$ and $\hat z$ are the pressure,
kinematic viscosity, volume
expansion coefficient, acceleration due to gravity and a unit
vector in the upward direction respectively. The
appearance of $T$ in the equation for $\B.u$ is crucial,
and changes the scaling exponents of $\B.u$. When the conditions
are right it may even change the scaling exponents from Kolmogorov 
to Bolgiano (up to anomalies) \cite{71MY}.

The second example is that of magneto-hydrodynamics (MHD), in which the
magnetic field $\B.b(\B.r,t)$ is driving the velocity field $\B.u(\B.r,t)$
according to \cite{Zeldovich}
\begin{eqnarray}
\frac{\partial \B.u}{\partial t}+\B.u\cdot
\B.\nabla \B.u&=&-\B.\nabla p +\B.b\cdot\B.\nabla \B.b +\nu \nabla^2 \B.u
 \ , \nonumber\\ 
\frac{\partial \B.b}{\partial t}+\B.u\cdot
\B.\nabla \B.b&=& \B.b\cdot\B.\nabla \B.u +\kappa \nabla^2 \B.b+\B.f \ .
\label{MHD}
\end{eqnarray}
These equations of motion conserve (in the inviscid, unforced limit)
three quadratic invariants, i.~e. the energy, magnetic helicity and cross
helicity \cite{93Bis}

Our main interest is in the properties of the statistical
objects characterizing the active fields, including their anomalous
scaling. Here ``anomalous scaling'' means that
multi-point correlation functions are homogeneous functions of their
arguments, with exponents that cannot be guessed from dimensional
analysis. Thus for example the field $\phi(\B.r,t)$ (with $\phi$ being $T$
or
$\B.b$ respectively) has simultaneous multi-point correlation functions
\begin{equation}
F^{(m)}(\B.r_1,\B.r_2,\!\cdots\!, \B.r_m)\equiv
\langle \phi(\B.r_1,t)\phi(\B.r_2,t)\!\cdots\!
\phi(\B.r_m,t)\rangle_f \ , \label{corr}
\end{equation}
where pointed brackets with subscript $f$ refer to
averaging over the statistics
of the advecting velocity field {\em and} of the forcing. The forcing
is taken to be white random noise with zero mean. When the
forcing is stationary in time this object is {\em time independent}.
Anomalous scaling means that
\begin{equation}
F^{(m)}(\lambda \B.r_1, \cdots, \lambda\B.r_m)
=\lambda^{\zeta_m} F^{(m)}(\B.r_1,\cdots, \B.r_m) \ , \label{zeta}
\end{equation}
with $\zeta_m$ having a non-trivial dependence on $m$. In what
follows we will assume that the advecting velocity field
itself is fully turbulent, and that its correlation functions are also
exhibiting scaling behavior like Eq.~(\ref{zeta}).

The main point of this paper is that the statistical
theory of the active fields calls for consideration of 
auxiliary passive fields that satisfy the same equations 
of motion as the active fields, but
{\em do not} affect the velocity field itself. In other words,
For the two problems at hand we consider the following equations of motion~:
\begin{eqnarray}
\frac{\partial \B.u}{\partial t}+\B.u\cdot
\B.\nabla \B.u&=&-\B.\nabla p+\nu \nabla^2 \B.u
+\alpha g T\hat z \ ,\nonumber\\
\frac{\partial T}{\partial t} + \B.u \cdot
\B.\nabla T &=& \kappa \nabla^2 T + f  \ ,\nonumber\\
\frac{\partial C}{\partial t}+\B.u\cdot
\B.\nabla C&=&\kappa \nabla^2 C+\tilde f \ , \label{passives}
\end{eqnarray}
on the one hand, and
\begin{eqnarray}
\frac{\partial \B.u}{\partial t}+\B.u\cdot
\B.\nabla \B.u&=&-\B.\nabla p +\B.b\cdot\B.\nabla \B.b +\nu \nabla^2 \B.u
 \ , \nonumber\\ 
\frac{\partial \B.b}{\partial t}+\B.u\cdot
\B.\nabla \B.b&=& \B.b\cdot\B.\nabla \B.u +\kappa \nabla^2 \B.b+\B.f\ ,
\nonumber\\
\frac{\partial \B.q}{\partial t}+\B.u\cdot
\B.\nabla \B.q&=& \B.q\cdot\B.\nabla \B.u +\kappa \nabla^2 \B.q+\tilde\B.f \
,
\label{passivev}
\end{eqnarray}
on the other.

Note that the velocity field that appears in the equations for
the passive fields is {\em the same} as the velocity field that results
from solving the coupled equations of the  associated equations for
the active fields. The forcing terms $\tilde f$ in Eq.~(\ref{passives})
(resp. $\tilde\B.f$ in Eq.~(\ref{passivev})) have the same statistics as the
forcing
terms $f$ in Eq. (\ref{active})  (resp. $\B.f$ in Eq. (\ref{MHD})), but they
must have
different realizations. While it is not true of course that the statistics
of the passive fields are independent of the statistics of the velocity 
fields, it is 
true that the statistics of the velocity fields  are independent of the
statistics of the passive forcing terms. This is, however, not the case
with the active forcing terms since these forcing terms affect the
active fields that affect in their turn the velocity fields. It is thus
not at all evident at first sight that there should be any relation,
apriori,
between the statistics of the active fields and their passive
counterparts. On the other hand, if there were such a relationship,
this would be very advantageous, since the statistics of the
passive fields is understood as explained next.

To understand the progress made in the context of passive fields
\cite{01FGV,01CV}, note
that the passive fields satisfy a linear equation of motion that
can be written as
\begin{equation}
\frac{\partial \phi(\B.r,t)}{\partial t}=\C.L \phi(\B.r,t)+f(\B.r,t) \ ,
\label{defL}
\end{equation}
with the actual form of the operator $\C.L$ determined by the problem at
hand.
In recent work \cite{01ABCPV,01CGP} it was clarified why and how passive
fields
exhibit anomalous scaling, when the velocity field is a generic
turbulent field. The key is to consider a
problem  associated with Eq. (\ref{defL}) which is the {\em decaying
problem} in which the forcing $f(\B.r,t)$ is put to zero.
The problem becomes then a linear initial value problem,
\begin{equation}
\partial \phi/\partial t =\C.L \phi\ , \label{decay}
\end{equation}
with a formal solution
\begin{equation}
\phi(\B.r,t) = \int d\B.r' \B.R(\B.r,\B.r',t) \phi(\B.r',0) \ , \label{oper}
\end{equation} 
with the operator 
\begin{equation}
\B.R\equiv T^+ \exp[{\int_0^t ds \C.L(s)}]\ , \label{defR}
\end{equation}
and $T^+$ being the time ordering operator.
Define next the {\em time dependent} correlation
functions of the decaying problem:
\begin{equation}
G^{(m)}(\B.r_1,\cdots ,\B.r_m,t)\equiv
\langle \phi(\B.r_1,t)\cdots
\phi(\B.r_m,t)\rangle \ . \label{defG}
\end{equation}
Here pointed brackets without subscript $f$ refer to the decaying
object in which averaging is taken with respect to realizations of
the velocity field only. As a result of Eq. (\ref{oper}) the decaying
correlation functions are developed by a propagator
$\C.P^{(m)}_{\underline{\B.r}|\underline{\B.\rho}}$, (with
$\underline {\B.r}\equiv \B.r_1,\B.r_2,\!\cdots\!,\B.r_m$)~:
\begin{equation}
G^{(m)}(\B.r_1,\!\cdots\!, \B.r_m,t)=\int\!\! d\underline{\B.\rho}
 \C.P^{(m)}_{\underline{\B.r}|\underline{\B.\rho}}(t)~
G^{(m)}(\B.\rho_1,\!\cdots\!, \B.\rho_m,0) \ . \label{defprop}
\end{equation}

In writing this equation we made explicit use of the fact that
the {\em initial} distribution of the passive field $\phi(\B.r,0)$
is statistically independent of the advecting velocity field. Thus
the operator $\C.P^{(m)}_{\underline{\B.r}|\underline{\B.\rho}}$
can be written explicitly
\begin{equation}
\C.P^{(m)}_{\underline{\B.r}|\underline{\B.\rho}}(t)\equiv
\langle  \B.R(\B.r_1,\B.\rho_1,t) \B.R(\B.r_2,\B.\rho_2,t)\cdots
\B.R(\B.r_m,\B.\rho_m,t)\rangle \ .
\end{equation}

The key finding \cite{01ABCPV,01CGP} is that the operator
$\C.P^{(m)}_{\underline{\B.r}|
\underline{\B.\rho}}$ possesses {\em left} eigenfunctions of eigenvalue
1, i.~e. there exist time-independent functions
$Z^{(m)}(\B.r_1,\B.r_2,\cdots ,\B.r_m)$ satisfying
\begin{equation}
Z^{(m)}(\B.r_1,\cdots, \B.r_m)=\int d\underline {\B.\rho}
\C.P^{(m)}_{\underline{\B.\rho}|\underline{\B.r}}(t)
Z^{(m)}(\B.\rho_1,\cdots, \B.\rho_m) \ .\label{zeromode}
\end{equation}
The functions $Z^{(m)}$ are referred to as ``Statistically Preserved
Structures'', being invariant to the dynamics, even though
{\em the operator is strongly time dependent and decaying}. How to form,
from these
functions, infinitely many conserved variables in the decaying
problem was shown in \cite{01ABCPV}, and is discussed again in
Sect. \ref{anal}. The functions
$Z^{(m)}(\underline{\B.r})$ are homogeneous functions of
their arguments, with anomalous scaling exponents $\zeta_m$:
\begin{equation}
Z^{(m)}(\lambda\underline{ \B.r}) = \lambda^{\zeta_m}
Z^{(m)}(\underline{\B.r})+\dots
\end{equation}
where ``$\dots$'' stand for subleading scaling terms. Since
Eq.~(\ref{zeromode})
contains $Z^{(m)}(\underline{\B.r})$ on both sides, the scaling exponent
$\zeta_m$
cannot be determined
from dimensional considerations, and it can be anomalous.
More importantly,
it was shown that the correlation functions of the forced case,
$F^{(m)}(\underline{\B.r})$ Eq. (\ref{corr}), have exactly the same scaling
exponents
as $Z^{(m)}(\underline{\B.r})$ \cite{01CGP}. In the scaling sense
\begin{equation}
F^{(m)}(\underline{\B.r})\sim Z^{(m)}(\underline{\B.r}) \ .
\label{scalesame}
\end{equation}
This is how anomalous scaling in passive fields is understood.
Lastly, we note that for the operator governing the time derivative
of Eq. (\ref{defG}), $Z^{(m)}(\underline{\B.r})$ is a zero mode. We
will use the terms ``Statistically Preserved Structures" and ``zero modes"
interchangeably.

Of course, returning to the active fields, it makes no sense to consider the
decaying
problem; as the active field decays, the statistics of the velocity field
changes, and there is very little to say. On the other
hand, we propose that it is possible to learn a great deal from considering
the forced
solutions, comparing the forced correlation functions of the active field
with those of the passive field when advected
by the same velocity field \cite{01CMMV}. The rest of this paper is devoted
to making
this point clear and solid.

In Sect. \ref{convection} we discuss the active problem (\ref{Boussinesq})
in comparison with the passive problem (\ref{passives}). A preliminary
report of the correspondence between these problems was presented in
\cite{02CCGP}. Since we are interested in points of principle rather than
quantitative details, we opt to work with a shell model of the turbulent
convection problem. We
will argue (cf. Sect. \ref{summary}) that there are excellent reasons to
believe that the
results found for  the shell model translate verbatim to the partial
differential
equations. The main result of Sect. \ref{convection} is that the forced
$2m$th-order
correlation functions of the active and passive fields are both dominated by
the
Statistically Preserved Structures of the decaying passive problem,
i.~e. the functions  $Z^{(2m)}(\underline{\B.r})$ of Eq. (\ref{zeromode}).
The anomalous
scaling exponents are the same for the passive and active {\em forced}
correlation
functions, they are universal (independent of the forcing $f(\B.r,t)$) and
determined by
the scaling exponents of  $Z^{(2m)}(\underline{\B.r})$. We present a careful
discussion of the role of the statistical correlations between the forcing
and the velocity field that exist in the active case, but are absent
in the passive case. In the present problem the net result of these
correlations is just an amplitude factor relating the moments of the
two fields. In Sect. \ref{vector} we turn to a shell model of
magneto-hydrodynamics.
On the face of it, this is a counter-example to the previous case: the
active and
passive fields exhibit radically different scaling exponents. The main
result
of this section is that nevertheless the Statistically Preserved Structures
of the passive problem are shown to dominate the statistics of the active
problem, but the existence of a conservation law in the latter results
in an exact cancellation of the amplitude of the leading zero mode.
We identify analytically the leading and subleading exponents of the
passive problem, and then observe the cancellation of the leading
contribution by the dynamics. The Summary section \ref{summary} presents
the general lesson for the statistical physics of the (nonlinear)
active problem. We propose that the zero modes of the auxiliary 
passive fields will always have a dominant role in the
statistics of active fields. The active fields will thus
share the same scaling exponents as the passive fields unless there
exist additional conservation laws for the active fields. In all cases
the calculation of the active scaling exponents can be achieved in the
context of the passive problem, which boils down to finding the 
zero modes of a linear operator. 
\section{Active and passive scalars in a model of turbulent convection}
\label{convection}
\subsection{Model and numerical results}
In this section we 
examine in detail a shell model of active and passive scalars
for which the statistical object can be computed to high
accuracy. We consider a model that reproduces the conservation laws and
the form of coupling between the active field and the velocity field in Eqs.
(\ref{active}) and (\ref{Boussinesq}). Our model is a variant
of the shell model studied in ref. \cite{Brand}:
\begin{widetext}
\begin{eqnarray}
\frac{\partial u_n}{\partial t}&=&ak_n(u_{n-1}^2-\lambda u_nu_{n+1})
+bk_n(u_nu_{n-1}-\lambda u^2_{n+1})-\nu k_n^2 u_n+T_n\ ,\label{convu}\\
\frac{\partial T_n}{\partial t}&=&\tilde ak_n(u_{n-1}T_{n-1}-\lambda
u_nT_{n+1})
+\tilde b k_n(u_nT_{n-1}-\lambda u_{n+1}T_{n+1})-\kappa k_n^2
T_n+f_0\delta_{n,0}\
,\label{convT}\\
\frac{\partial C_n}{\partial t}&=&\tilde ak_n(u_{n-1}C_{n-1}-\lambda
u_nC_{n+1})
+\tilde bk_n(u_n C_{n-1}-\lambda u_{n+1}C_{n+1})-\kappa k_n^2
C_n+f_0\delta_{n,0} \ .
\label{convC}
\end{eqnarray}
\end{widetext}
In this model all the field variables are real and $n$ stands for the index
of a 
shell of wavevector
$k_n=k_0\lambda^n$, with $n=0,1,\cdots, N-1$. We take $\lambda=2$, and the
parameters used
in the simulation are $a=0.01$, $\tilde a=\tilde b=b=1$,
$k_0=1$, $\kappa=\nu=5\times 10^{-4}$. The number of shells
is $N=30$, and the forcing is white noise of zero mean on the first shell.

\begin{figure}
\centering
\includegraphics[width=.45\textwidth]{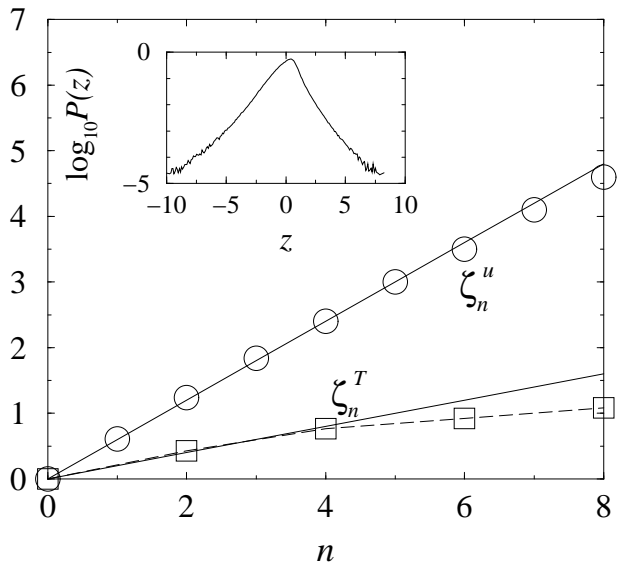}
\caption{The scaling exponents $\zeta_n^u$ of the velocity field (circles)
and $\zeta_n^T$ of the active scalar field (squares) for even $n$'s.
The solid lines are respectively $3n/5$ and $n/5$ for the velocity
and the active scalar fields. Shown in the inset is
the PDF of $z\equiv u_n/\langle  u_n^2 \rangle^{1/2}$
at shell $n=14$.}
\label{scexp}
\end{figure}

Without the coupling to $T_n$, the velocity equation has
an inviscid unstable Kolmogorov fixed point, $u_n\sim k_n^{-1/3}$.
This is changed by the coupling \cite{Brand}, and the system of equations
for $T_n$ and $u_n$ exhibits an inviscid unstable Bolgiano
fixed point, $u_n\sim k_n^{-3/5}$, $T_n\sim k_n^{-1/5}$.
The chaotic dynamics renders the statistics of the velocity field strongly
non-Gaussian (cf. inset in Fig. \ref{scexp}). The exponents $\zeta_n^T$
for the active scalar are markedly anomalous, whereas, for the velocity,
they appear closer to normal (see Fig. \ref{scexp}).  The equation of motion
for the passive field $C$ is identical to the equation of motion of $T$,
but it does not affect the velocity field $u$. This equation has
a $C\to -C$ symmetry, whereas the coupled system of $T$ and $u$ lacks
this symmetry. This difference is reflected in the statistics of the two
fields.

\begin{figure}
\centering
\includegraphics[width=.45\textwidth]{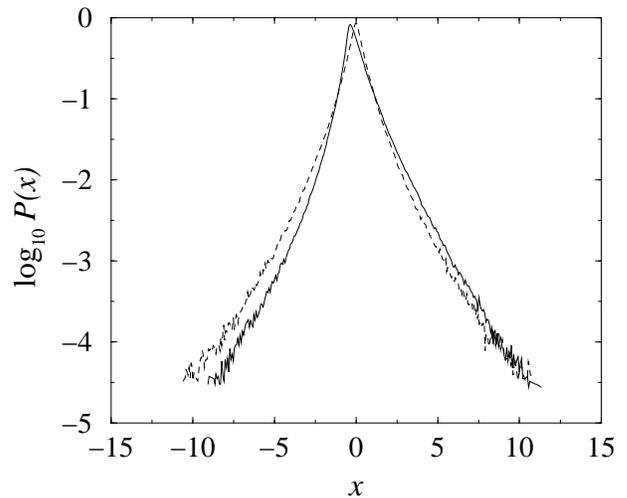}
\caption{The PDF's of the active (solid) and passive (dashed)
scalars at shell $n=14$. Note that the PDF of
the active scalar is asymmetric.}
\label{PDF}
\end{figure}

\begin{figure}
\centering
\includegraphics[width=.45\textwidth]{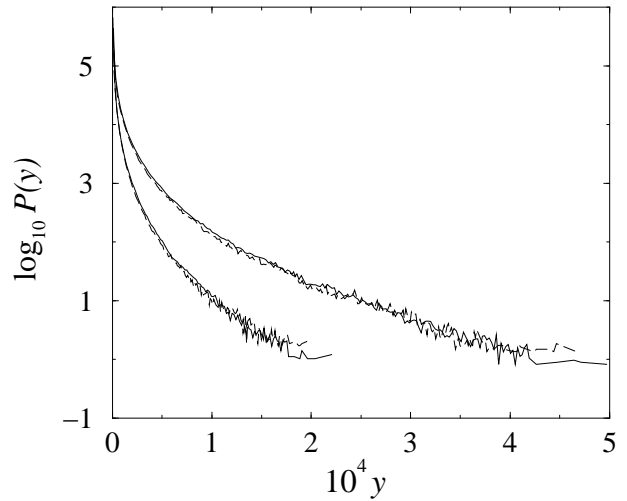}
\caption{
The PDF's of $y$ where $y=\tilde T_n^2$ (solid)
or $\beta C_n^2$ (dashed) at shells $n=9$ and 14.}
\label{PDFsq}
\end{figure}

To demonstrate this
difference between the active and passive fields, we show in Fig. \ref{PDF}
the probability distribution functions (PDF) of $x= \phi_n/\langle
\phi_n^2\rangle^{1/2}$  where $ \phi_n$ is $ T_n$ or $C_n$,
for $n=14$. One clearly sees the symmetry of the PDF
of the passive scalar, in contradistinction to the asymmetry
of the PDF of the active scalar. This is typical to all $n$
in the inertial range. This is a demonstration of the
discussion after Eq. (\ref{phioft}). For the passive scalar the
odd moments vanish, whereas for the active scalar they all exist.
The situation is altogether different for the statistics of
even moments. To demonstrate the difference we plot in Fig. \ref{PDFsq}
the (typical) PDF of $\tilde T_n^2$ and $C_n^2$ for $n=9$ and 14,
where $\tilde T_n\equiv T_n-\langle T_n\rangle$. In plotting
we realize that the passive scalar is defined up to a constant,
so for the passive scalar the PDF is plotted for the rescaled
variable $\beta C^2_n$, where
\begin{equation}\beta=\langle \tilde T _n^2\rangle_f /
\langle C_n^2\rangle_f \approx 0.6327 \ . \label{defbeta}
\end{equation} 
Note that there is only one
numerical freedom $\beta$, constant for all $n$ in the inertial range.
An understanding of this numerical constant based on dynamical
considerations
is given in the next subsection.

\begin{figure}
\centering
\includegraphics[width=.45\textwidth]{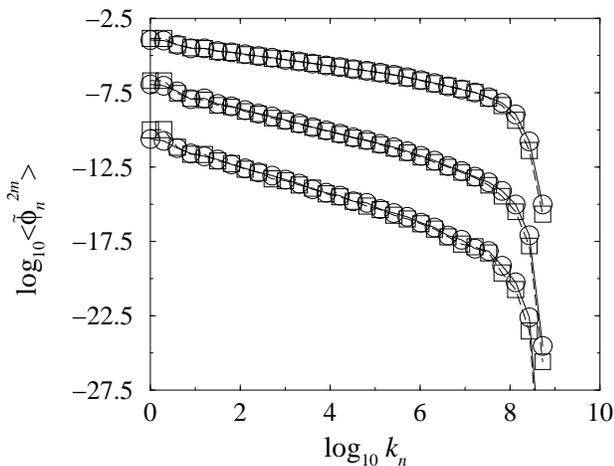}
\caption{
The even-order structure functions $\langle \tilde T_n^{2m} \rangle_f$
(circles)
and $\langle \beta^m C_n^{2m} \rangle_f$ (squares), with
$m=1$, 2 and 3, from top to bottom.}
\label{structure}
\end{figure}

\begin{figure}
\centering
\includegraphics[width=.45\textwidth]{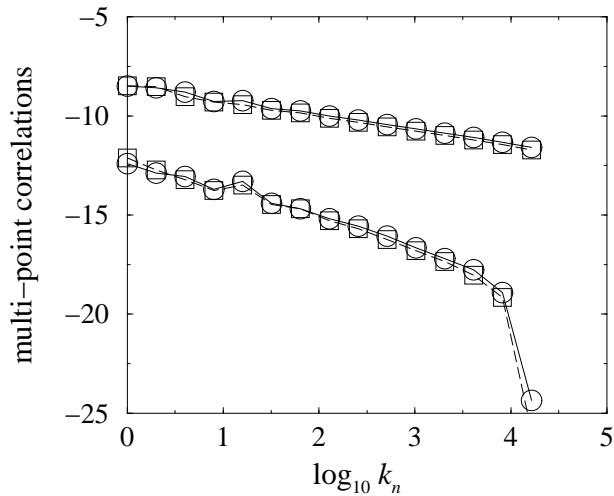}
\caption{Upper:
$\log_{10}\langle \tilde T_n^2\tilde T_{n+5}^2 \rangle_f$ (circles) and
$\log_{10}\langle \beta^2 C_n^2C_{n+5}^2 \rangle_f$ (squares). Lower:
$\log_{10} \langle \tilde T_n^2\tilde T_{n+5}^2\tilde T_{2n}^2 \rangle_f$
(circles) 
and $\log_{10} \langle \beta^3C_n^2C_{n+5}^2C_{2n}^2 \rangle_f$ (squares).}
\label{multi}
\end{figure}

We find very close agreement of all the PDF's in the
inertial range.
The identity of the PDF's of $\tilde T_n^2$ and $C_n^2$ translates
automatically to the identity of the even-order
structure functions $F^{(2m)}(k_n)\equiv \langle \tilde
\phi_n^{2m}\rangle_f$,
where $\tilde \phi_n = \tilde T_n$ or $C_n$ (up to a constant $\beta^m$).
This is demonstrated in Fig. \ref{structure}.
We see that the 2nd, 4th and 6th-order
structure functions are barely distinguishable, with the
same scaling exponents in the inertial range.
Finally, we demonstrate that the identity of the statistics
of the squares of the passive and
active scalars transcends structure functions. Consider
for example the multi-point correlation functions
 $\langle \tilde T_n^2
\tilde T^2_{n+5}\rangle$
and $\langle \tilde T^2_n \tilde T^2_{n+5}\tilde T^2_{2n}\rangle$.
In Fig. \ref{multi} these  correlation functions
are compared to their passive counterparts.
The conclusion is that again the multi-point correlation
functions are indistinguishable once the passive
ones are rescaled by $\beta^q$ where $q$ is the overall
order of the correlation function.
\subsection{Analysis of the results}
\label{anal}
\begin{figure}
\centering
\includegraphics[width=.45\textwidth]{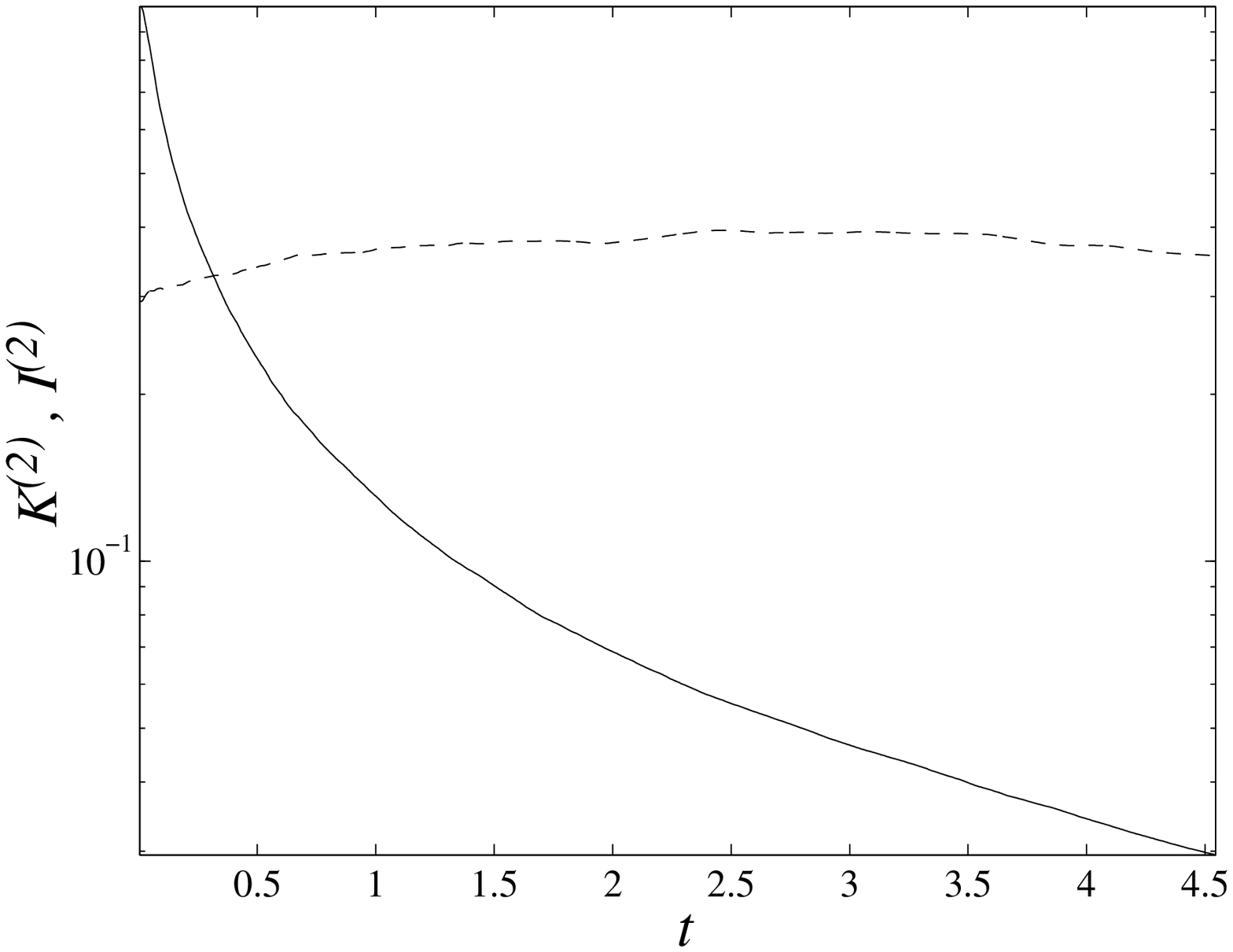}
\centering
\includegraphics[width=.45\textwidth]{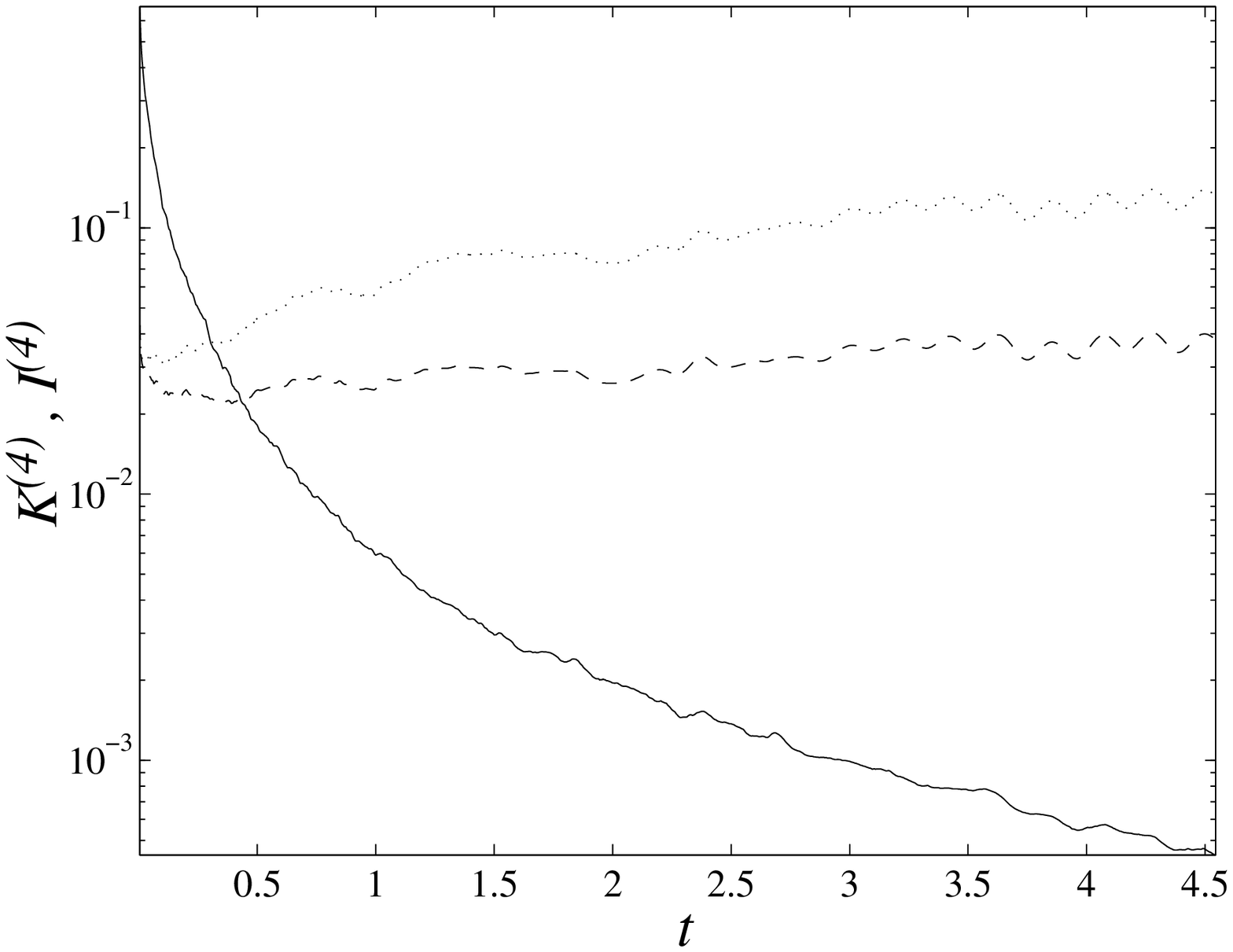}
\centering
\includegraphics[width=.45\textwidth]{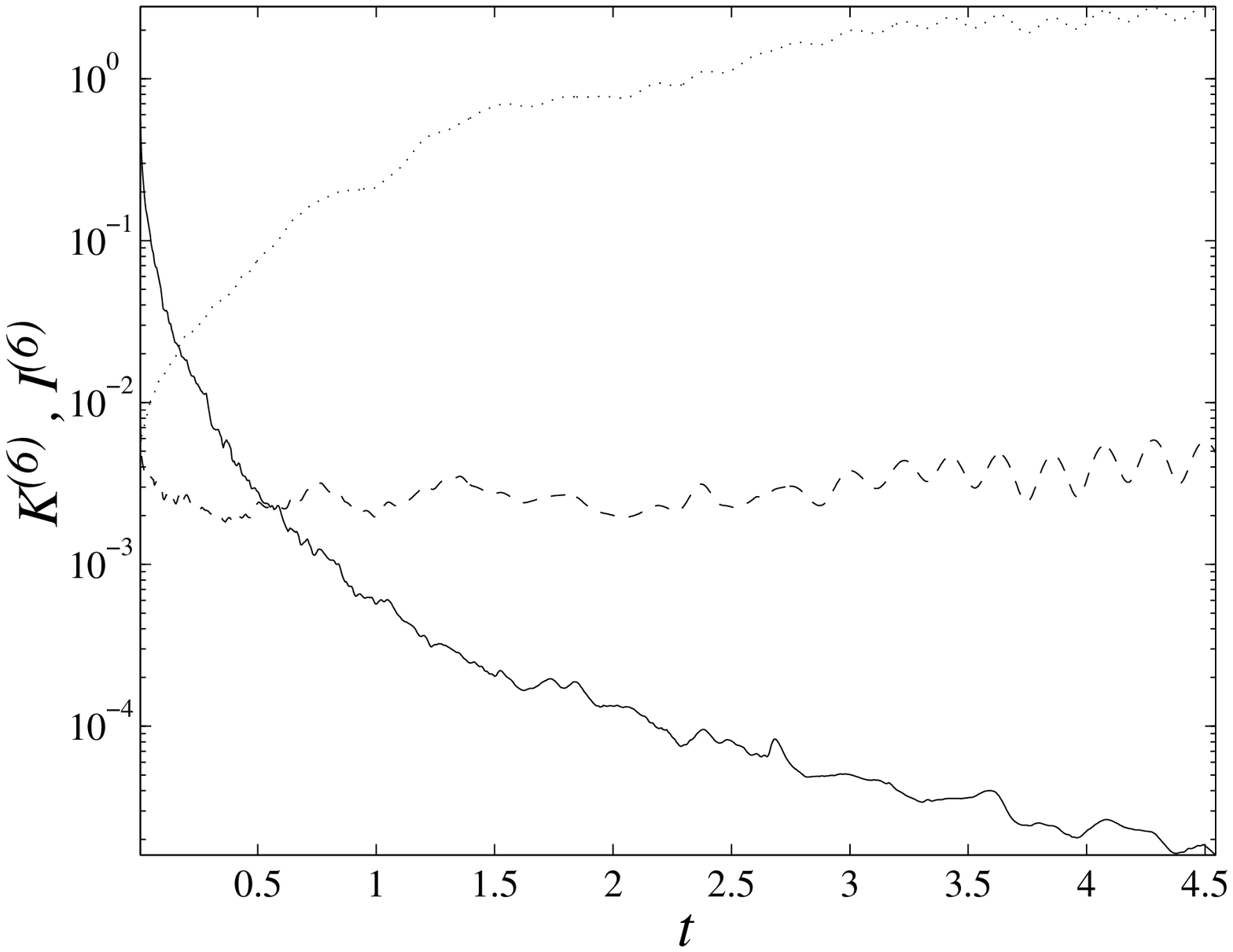}
\caption{The decaying objects $K^{(2m)}$ (solid lines) and the conserved
objects
$I^{(2m)}$ (dashed lines) as a function of time, for $m=1,2$ and 3. Time is
measured
here in units of the largest scale eddy turn over time $\tau_0\equiv
[k_0 \sqrt{S_2(k_0)}]^{-1}\approx 22$. In
panels b and c we include in dotted lines the quantity $I^{(2m)}$ in which
we replaced
$F^{(2m)}$ by its dimensional prediction $[F^{(2)}]^m$. We see that using
the
dimensional exponent does not make $I^{(2m)}$ time invariant.}
\label{SPSpas}
\end{figure}

To understand the results we start with the passive field,
demonstrating that its forced structure functions are
actually Statistically Preserved Structures. Consider then
the decaying passive problem, i.~e. Eq.~(\ref{convC}) without the
forcing term. The initial value problem for the time dependent
structure functions $G^{(2m)}(k_n,t)\equiv \langle C_n^{2m}(t) \rangle$
is the shell analog of Eq. (\ref{defprop}),
\begin{equation}
G^{(2m)}(k_n,t) =  \C.P^{(2m)}_{n,n'}(t) G^{(2m)}(k_{n'},t=0) \ ,
\label{defprop2}
\end{equation}
which defines the $2m$th-order
propagator $\C.P^{(2m)}_{n,n'}(t)$. Here and below, repeated indices are
being summed over. In Fig. \ref{SPSpas}
we show typical decay plots for the quantity $K^{(2m)}\equiv \sum_n
G^{(2m)}(k_n,t)$ for
$m=1,2,3$, starting from the initial conditions
$G^{(2m)}(k_n,t=0)=\delta_{n,16}$.

\begin{figure}
\centering
\includegraphics[width=.45\textwidth]{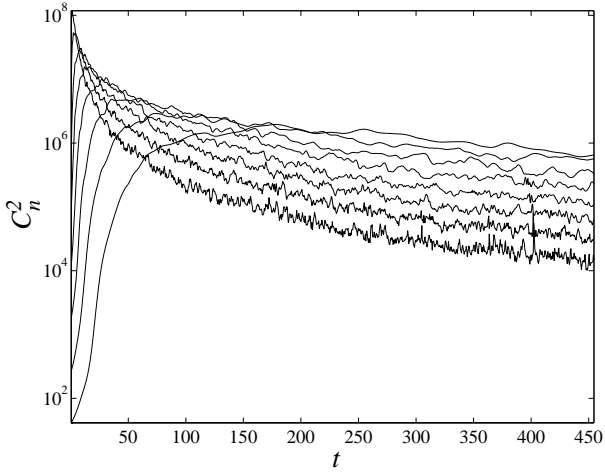}
\caption{An example of the fate of an initial value term as a function of
time,
in units of $\tau_0$. The initial amplitudes are inverted in order compared
to the 2nd order zero modes of the passive field. Shown are shells
$n=5,\:7,\dots,\:19$. }
\label{initsc}
\end{figure}

Statistically Preserved Structures in this case 
represent left-eigenfunctions
$Z^{(2m)}(k_n)$ of eigenvalue 1 satisfying
\begin{equation}
Z^{(2m)}(k_{n'}) =  Z^{(2m)}(k_n)\C.P^{(2m)}_{n,n'}(t)  \ . \label{defZ}
\end{equation}
The statement that we want to demonstrate is that the forced
structure functions $F_n^{(2m)}$ of the passive scalar scale like these
eigen-modes of the decaying problem:
\begin{equation}
F^{(2m)}(k_n) \equiv \langle C^{2m}_n\rangle_f \sim Z^{(2m)}(k_n) \ .
\label{true}
\end{equation}
To demonstrate this we use the method of \cite{01ABCPV} and
define the quantities $I^{(2m)}$,
\begin{equation}
I^{(2m)} = \sum_n G^{(2m)}(k_n,t) F^{(2m)}(k_n) \ .
\end{equation}
Using Eqs. (\ref{defprop2}) and (\ref{defZ}) we see that if
Eq.~(\ref{true}) is obeyed, than the quantities $I^{(2m)}$ are time
independent. Indeed, in Fig. \ref{SPSpas} we demonstrate the stationarity
of these objects, thus supporting Eq. (\ref{true}). The analytic
explanation as to why the forced solutions agree with the Statistically
Preserved Structures of the decaying problem was provided in \cite{01CGP}.
Before turning to the active field, it is worthwhile to observe
how any initial condition of the decaying passive field lands
on the scaling solution that is represented by the Statistically
Preserved Structure. Consider the initial value experiment that is
reported in Fig. \ref{initsc}.  Here we start,
as an example, from the initial value
$C_n(t=0)\propto k_n^{2/3}$. In this initial condition the order
of the amplitudes is inverted with respect to the spectrum of
the passive scalar.  We plot, as a function of time,
the trajectories of
$C_n(t)$ as computed just from this initial condition, averaged over
650 realizations. We see that the trajectories
land on a decaying scaling solution in which the order of the amplitudes and
the ratios between them are identical to the spectrum of the zero mode of
the 
passive field; the decay that we see, at a rate proportional to $1/t^2$,
is entirely due to dissipative effects, as explained in some detail
in \cite{01CGP}.

Finally, we need to understand how the forced active scalar $T$ falls on
the Statistically Preserved Structure of the decaying passive
problem, and what is the origin of the factor $\beta$ in (\ref{defbeta}).
To this aim we note that both equations for passive and active fields can
be written as
\begin{equation}
\frac{\partial \phi_n}{\partial t}=\C. L_{n,n'}\phi_{n'}
+f_0\delta_{n,0}\ . \label{convphi}
\end{equation}
This equation has a formal solution in the form
\begin{equation}
\phi_n(t) = R_{n,n'}(t|0)\phi_{n'}(t=0)
+\int_0^t d\tau R_{n,0}(t|\tau)f_0(\tau)\ , 
\end{equation}
where $R_{n,n'}(t|\tau)$ is the shell analog of the operator (\ref{defR}),
\begin{equation}
R_{n,n'}(t|\tau) \equiv \left\{T^+ \int_\tau^t \exp
[\C.L(s)]ds\right\}_{n,n'} \ .
\label{phioft}
\end{equation}
The first difference between the
active and passive fields is encountered when we take the
average of this equation. For the passive case, the average
can be taken by decorrelating $f_0$ and $R_{n,n'}$. 
Since the mean of the force $f_0$ vanishes, we get
\begin{equation}
\langle C_n \rangle_f=0  
\label{meanC}
\end{equation}
Such a decorrelation is, however, not 
allowed in the active case since the forcing $f_0$
is correlated with $T$, which is itself correlated 
with $\B.u$ and thus with $R_{n,n'}$. Hence 
\begin{equation}
\langle T_n \rangle_f=\langle R_{n,n'}(t|0)T_{n'}(t=0)\rangle_f
\nonumber\\+\int_0^t d\tau \langle R_{n,0}(t|\tau)f_0(\tau)\rangle_f 
\label{meanT}
\end{equation}
and $T$ has a non-zero mean.
Similarly, the passive scalar has zero odd moments and its PDF is
symmetric. On the other hand, the active scalar has nonvanishing odd
moments and its PDF is asymmetric.

In spite of this great difference between the active and passive scalars,
there is a close affinity between the active field and
the Statistically Preserved Structures of the passive field. 
To see this, we note that
the first term on the RHS of Eq. (\ref{meanT}) represents a decaying field.
We expect that if the initial condition $T_n(t=0)$ has any component on
the Statistically Preserved Structure of the passive field, it will quickly
relax everything else and will land exactly on that solution. In this
respect it is just the same as the initial value experiment reported
in Fig. \ref{initsc}. In terms of the relative amplitudes of
the different $n$ shells
there is nothing in the fate of the initial value term to distinguish the
active and the
passive fields.
The second term on the RHS of Eq.~(\ref{meanT}) is more subtle. First, we
note
that for every value of $\tau$ we again face a decaying experiment that
takes
place between the times $\tau$ and $t$. In the language of the passive
field, the integrand can be read from a decaying field with initial
condition $C_n(t=\tau)=f_0(\tau)\delta_{n,0}$. Indeed, in our simulations
below
this is precisely how we evaluate integrals of this type. We break the
interval $[0,t]$ into $N$ sub-intervals $\{\tau_i\}_{i=1}^N$, $\tau_i =
(i/N)t$,
and start a decaying experiment with
initial conditions $f(\tau_i)\delta_{n,0}$. Measuring $C_n(t)$ and summing
up
all the contributions yields an approximation to the integral. Every term in
the integrand is expected to land, for most of the time $t-\tau$, on the
scaling solution of the passive field, in much of the same way that the
initial value term does.
Thus both terms in Eq. (\ref{meanT}) are expected to scale like the passive
field, which nevertheless itself has zero amplitude due to the symmetry.

\begin{figure}
\centering
\includegraphics[width=.45\textwidth]{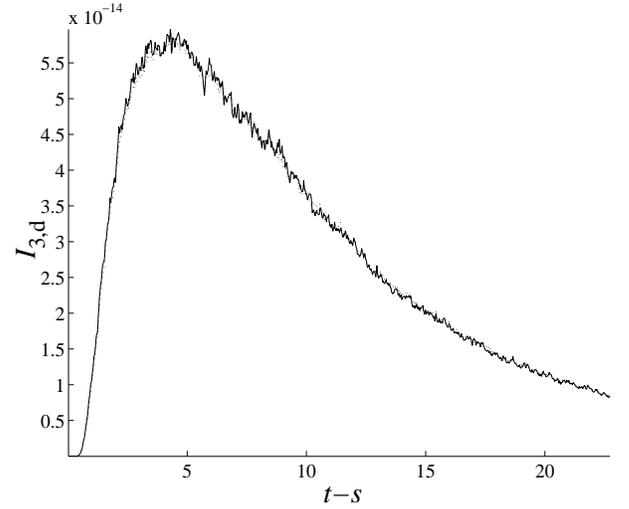}
\caption{Comparison of the integrands in Eqs (\ref{one}) (the passive case)
and (\ref{two}) (the active case) for $n=10$. The plots are indistinguishable.}
\label{diagonal}
\end{figure}

The correlation effects that play a role for the active scalar will be
responsible for the factor $\beta$ that we discovered in the numerics.
To see this we need to consider the 2nd order structure functions:
\begin{eqnarray}
&&F^{(2)}(k_n)= \langle [R_{n,n'}(t|0)\phi_{n'}(t=0)]^2\rangle_f\label{F2}\\
&&+2\langle R_{n,n'}(t|0)\phi_{n'}(t=0)\int_0^t d\tau
R_{n,0}(t|\tau)f_0(\tau)\rangle_f
\nonumber\\
&&+\int_0^t d\tau' d\tau''\label{second}\langle
R_{n,0}(t|\tau')R_{n,0}(t|\tau'')f_0(\tau')f_0(\tau'')\rangle_f \ .
\nonumber
\end{eqnarray}
For sufficiently long time the first two terms, denoted below as
$\C.I_1$ and $\C.I_2$ respectively, do not contribute to the
structure functions, and any difference between the active and passive
fields must be ascribed to the last term. The last term, denoted here
as $\C.I_3$, has a ``diagonal"
contribution, which is obtained for $\tau'=\tau''$ and an ``off-diagonal"
contribution, which is the rest of the integral for which  $\tau'\ne
\tau''$.
For the passive field, $R_{n,n'}$ and $f_0$ decouple and only the 
diagonal part exists. For the active field there is no decoupling. 
Denoting this term $\C.I_{3,d}$, it
reads respectively
\begin{eqnarray}
\!\!\!\!\C.I_{3,d}\!\!&=&\!\!\int ds\langle
R_{n,0}(t|s)R_{n,0}(t|s)\rangle_f f_0^2 \quad \text{(passive)}\ ,
\label{one}\\
\!\!\!\!\C.I_{3,d}\!\!&=&\!\!\!\int_0^t \!\!ds\langle
R_{n,0}(t|s)R_{n,0}(t|s)f_0(s)f_0(s)\rangle_f \ \text{(active)}. \label{two}
\end{eqnarray}
In Fig. \ref{diagonal} we compare the integrands of these two expressions,
measured directly in our simulation as explained above.
\begin{figure}
\centering
\includegraphics[width=.45\textwidth]{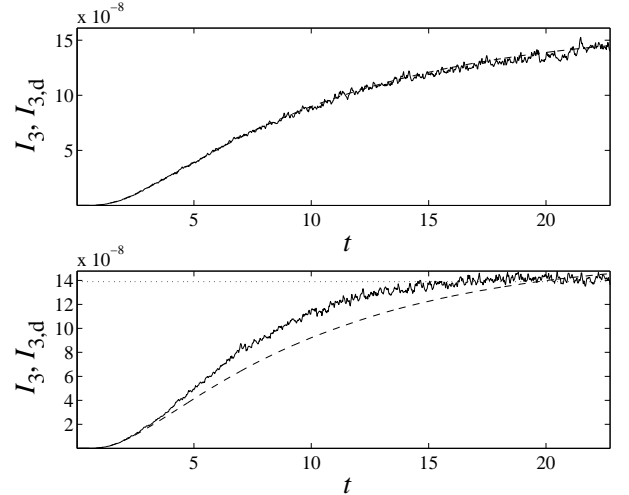}
\caption{The integral $\C.I_3$ in comparison to the diagonal
part $\C.I_{3,d}$ for $n=10$. Upper panel: the passive field. The integral agrees
with the diagonal part at all times. Lower panel: the active field. The
deviations are due to the non-vanishing contribution of the off-diagonal
integral, which is also displayed in the next figure. The dashed
line in both panels represents the stationary value of the
corresponding second order structure function.}
\label{intmindiag}
\end{figure}
\begin{figure}
\centering
\includegraphics[width=.45\textwidth]{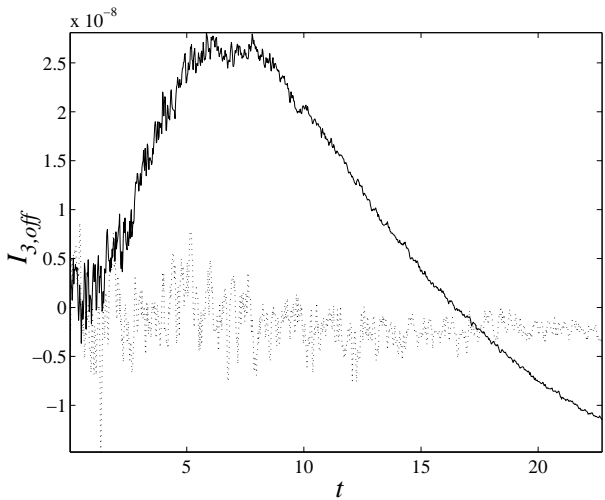}
\caption{The off-diagonal integral for $n=10$, computed as
$\langle [\phi_n(t)-R_{n,n'}(t|0)\phi_{n'}(t=0)]^2\rangle -\C.I_{3,d}$  for
the passive and active fields
respectively. For the passive field (dotted line) it fluctuates around zero,
while for
the active field it begins positive, and then turns negative. For longer
times it saturates at a constant negative value, giving rise to the
factor $\beta$.}
\label{beta}
\end{figure}
We can see that there is not much difference
between them; the diagonal term cannot be blamed for the factor $\beta$.
On the other hand, in Fig. \ref{intmindiag} we show the full integral
$\C.I_3$ and compare it with its diagonal term. We see that in the passive
case the diagonal part is everything, whereas in the active case there
is a difference. Lastly, we show that this difference is precisely
the source of the factor $\beta$. In Fig. \ref{beta} we show
$\langle [\phi_n(t)-R_{n,n'}(t|0)\phi_{n'}(t=0)]^2\rangle
-\C.I_{3,d}$ for the passive and the
active fields. The former fluctuates all the time around zero. The latter
is positive initially, and then becomes negative. For later times it
saturates
at a negative value that is precisely responsible for the factor $\beta$.

In summary, we find that the even correlation functions of the active
and passive scalars share the same scaling exponents simply because
the zero modes of the decaying passive problem dominate the statistics
of both fields. It is thus possible to understand the anomalous
statistics of the active field in the same way as that of the passive
field. We believe that this is a significant result that should be
put to further experimental and numerical tests in the PDE version
of the problem. We will return to this point in the discussion.
\section{Active and passive fields in a model of magnetohydrodynamics}
\label{vector}
\subsection{Model and Numerical Results}

In this section we examine a shell model that reproduces the type
of coupling and the conservation laws of Eqs.~(\ref{MHD}).  We need to be
careful about the dynamo effect which we want to avoid in order to
have stationary statistics. We thus construct the model to mimic
2-dimensional
MHD, in which there is an inverse cascade of energy. Accordingly we need to
have large scale damping terms in the velocity equation, and
a force at intermediate scales. In all our simulations below we force both
the velocity and the active
fields on shells 10 and 11 (denoted by $n_f$), using white noise of zero
mean. We run the
model with 35 shells. The equations are an adaptation of the MHD
shell model of \cite{98GC} to the Sabra shell model \cite{98LPPPV}. All
field variables are complex numbers:
\begin{widetext}
\begin{eqnarray}
\label{sabrau}
\frac{d u_n}{dt}&=& i k_n [a \lambda (u_{n+1}^*u_{n+2}-b_{n+1}^*b_{n+2})
+ b (u_{n-1}^*u_{n+1} - b_{n-1}^*b_{n+1})-c \lambda^{-1}(u_{n-2}u_{n-1} -
b_{n-2}b_{n-1})]\nonumber\\
&&+ f'_n \delta_{n,n_f} + \nu k_n^2 u_n+\tilde \nu k_n^{-4} u_n\ , \\
\frac{d b_n}{dt}&=& i k_n [\tilde{a} \lambda
(u_{n+1}^*b_{n+2}-b_{n+1}^*u_{n+2})
+ \tilde{b} (u_{n-1}^*b_{n+1} - b_{n-1}^*u_{n+1})
-\tilde{c} \lambda^{-1}(u_{n-2}b_{n-1} - b_{n-2}u_{n-1})]\nonumber\\
&&+ f_n \delta_{n,n_f} +\kappa k_n^2 b_n +\tilde \kappa k_n^{-4} b_n\ .
\label{sabrab}
\end{eqnarray}
\end{widetext}
The coefficients $a$, $b$, $c$, $\tilde{a}$, $\tilde{b}$ and $\tilde{c}$
can be parametrized as follows~:
\begin{equation}
\begin{array}{l@{\quad}l@{\quad}l}
a = 1\ ,&b = -\delta\ ,&c = -(1-\delta)\ ,\\
\tilde{a}=1-\delta-\delta_m\ ,&\tilde{b}=\delta_m\ ,& \tilde{c}=1-\delta_m\
.
\end{array}
\label{sabraparam}
\end{equation}
\label{sixparam}
This choice ensures the conservation of the total energy and ``cross
helicity"
in the inviscid limit $\nu=\tilde \nu=\kappa=\tilde \kappa = f = f' =0$,
\begin{eqnarray}
E&=&\frac{1}{2}\sum_n (|u_n|^2 + |b_n|^2)\ , \label{toten}\\
K&=&\sum_n \Re(u_n^* b_n)\ .\label{crosshe}
\end{eqnarray}
To mimic the magnetic helicity, we can write down a generalized quantity
\begin{equation}
H=\frac{1}{2}\sum_n \mathrm{sign}(\delta-1)^n\frac{|b_n|^2}{k_n^\alpha},
\label{magnhe}
\end{equation}
with $\alpha>0$ a fixed parameter. We demand conservation of this
generalized
``magnetic helicity", together with absence of dynamo effect. This implies
\begin{eqnarray}
\delta>1&\rightarrow&
\left\{\begin{array}{l}
\delta = 1 + \lambda^{-\alpha}\ ,\\
\delta_m=- 1/(\lambda^\alpha - 1)\ ,
\end{array}\right.
\label{deltagt1}
\end{eqnarray}
On the other hand, when $\delta<1$ one can have dynamo, and therefore
no stationary statistics.

In addition to the conservation laws the equations
of motion remain invariant to the phase transformations
$u_n\to u_n\exp(i\phi_n)$ and $b_n\to b_n\exp(i\psi_n)$. The conditions are
\begin{eqnarray}
\phi_n + \phi_{n+1} - \phi_{n+2} &=& 0\ ,\label{phases1}\\
\phi_{n} + \psi_{n+1} - \psi_{n+2} &=& 0\ ,\label{phases2}\\
\psi_n + \phi_{n+1}-\psi_{n+2} &=&0\ ,\label{phases3}\\
\psi_{n} + \psi_{n+1} - \phi_{n+2} &=& 0\ .\label{phases4}
\end{eqnarray}
This implies $\psi_n = \phi_n$ $\forall n$.

The passive field is denoted by $q_n$, whose evolution is given by an
equation similar to Eq. (\ref{sabrab}), i.~e.
\begin{widetext}
\begin{eqnarray}
&&\frac{d q_n}{dt}= i k_n [\tilde{a} \lambda
(u_{n+1}^*q_{n+2}-q_{n+1}^*u_{n+2})
+ \tilde{b} (u_{n-1}^*q_{n+1} - q_{n-1}^*u_{n+1})- \tilde{c}
\lambda^{-1}(u_{n-2}q_{n-1} - q_{n-2}u_{n-1})]\nonumber\\
&&+\kappa k_n^2 q_n +f_n\delta_{n,n_f}\ . \label{sabraq}
\end{eqnarray}
\end{widetext}
The fields $q_n$ and $b_n$ are advected by the same velocity field, however
$b_n$ is
active, while $q_n$ is passive. The inviscid passive equation has only
one conserved variable, i.~e. Eq. (\ref{magnhe}) with $q_n$ replacing
$b_n$. It also satisfies the same phase relations as the active
field. We want to know whether the
scaling properties of $b_n$ are determined once again by the Statistically
Preserved
Structures of the decaying problem of the passive field $q_n$.
\begin{figure}
\centering
\includegraphics[width=.45\textwidth]{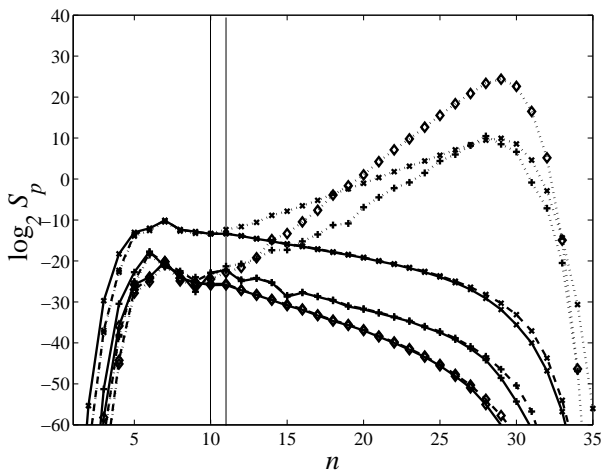}
\caption{Structure functions of order 2 ($\times$), 3 (+)
and 4 ($\diamond$) for the passive field (dotted line), active
field (dashed line) and velocity field (solid line). The two vertical
lines denote the forcing shells. Note
that the scaling exponents of the active field and the velocity field
coincide.
The parameters are $N=35$, $\alpha=2$, $k_0=0.0625$, $\nu=10^{-12}$,
$\tilde\nu
=10^{-3}$. The forcing is white noise on shells 10,11.}
\label{spectra}
\end{figure}

In Fig. \ref{spectra} we show the spectra of the passive and active fields
respectively, obtained from a direct numerical simulation with the
parameters as detailed in the figure legend. This appears like a
striking counter-example to the results of the previous section: the
two fields have totally different scaling behaviors. The active field
has ``standard" scaling exponents $\eta_p$, 
defined by $\langle |b_n|^p\rangle \sim k_n^{-\eta_p}$, that coincide
with those of the velocity field, defined by
$\langle |u_n|^p\rangle \sim k_n^{-\zeta_p}$,
and the spectrum decays like a power
law in the ``inertial range" which is between the forcing and the
dissipative scales. We estimate from the numerics $\eta_2=\zeta_2\approx .67$,
$\eta_3=\zeta_3\approx 1.0$ and $\eta_4=\zeta_4\approx 1.33$, 
in close correspondence
with the Kolmogorov dimensional predictions.  The passive
field has exponents, defined similarly by 
$\langle |q_n|^p\rangle \sim k_n^{-\beta_p}$,
that are with a different sign! Its spectrum is an increasing
function of
$k_n$ in the inertial range. We measure $\beta_2\approx -1.33$,
$\beta_3\approx -2$ and $\beta_4\approx -2.67$. If we assume that the
passive field
lands on the Statistically Preserved Structures of the passive decaying problem,
then it
appears that the active field does not do so.

In the rest of this section we will show that this is actually not a
counterexample
to the proposition that the active field lands on Statistically Preserved
Structures of the decaying passive field. It does. What happens here is
that,
due to the conservation law Eq. (\ref{crosshe}), the amplitude of the
leading
Statistically Preserved Structure with the negative scaling exponent
is exactly zero. The active
field then lands on a sub-leading zero mode, which has standard, positive
scaling exponent (the positive sign refers to $\B.r$-space representation,
as in Eq. (\ref{zeta})). 
\subsection{Analysis of the results}

\begin{figure}
\centering
\includegraphics[width=.45\textwidth]{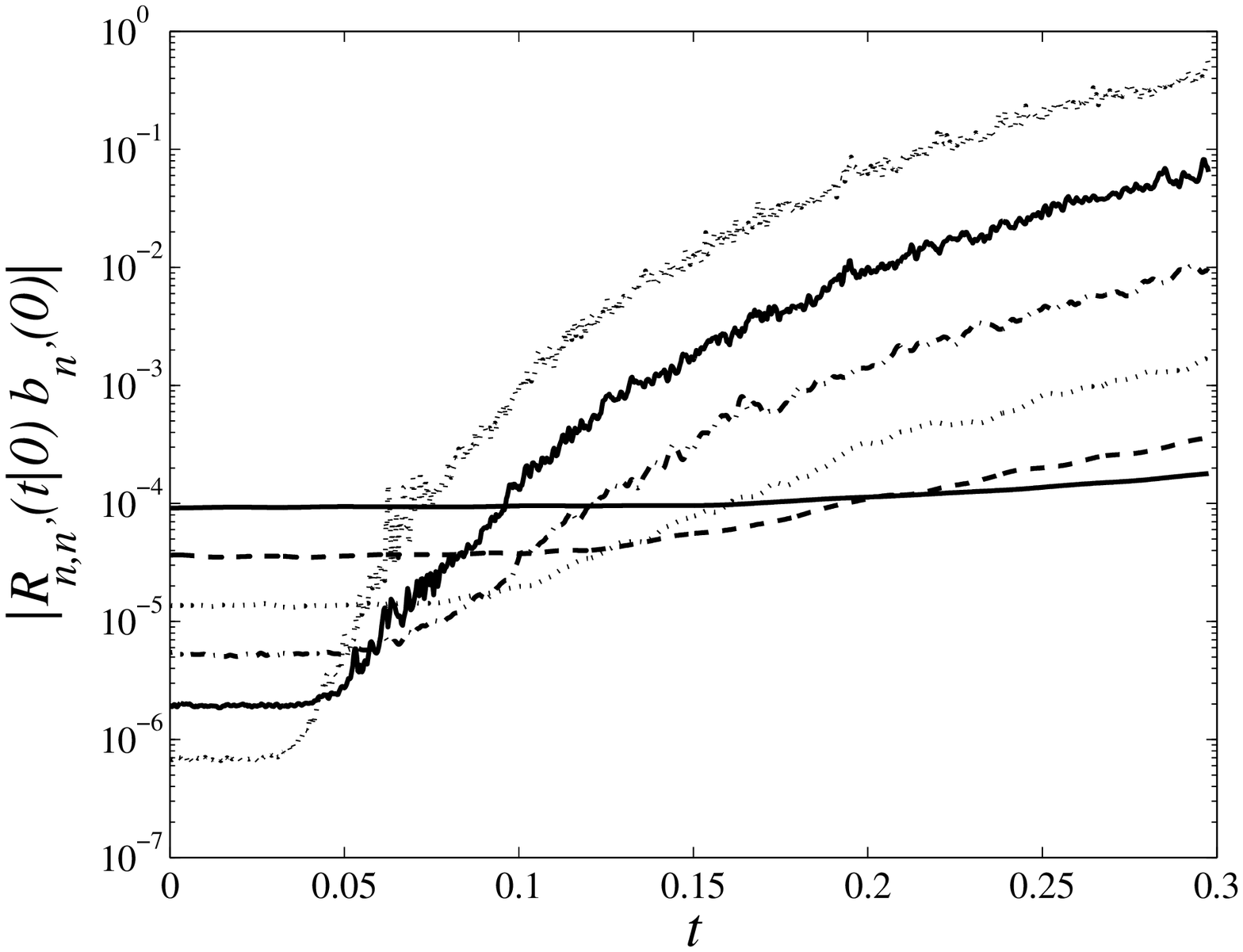}
\includegraphics[width=.45\textwidth]{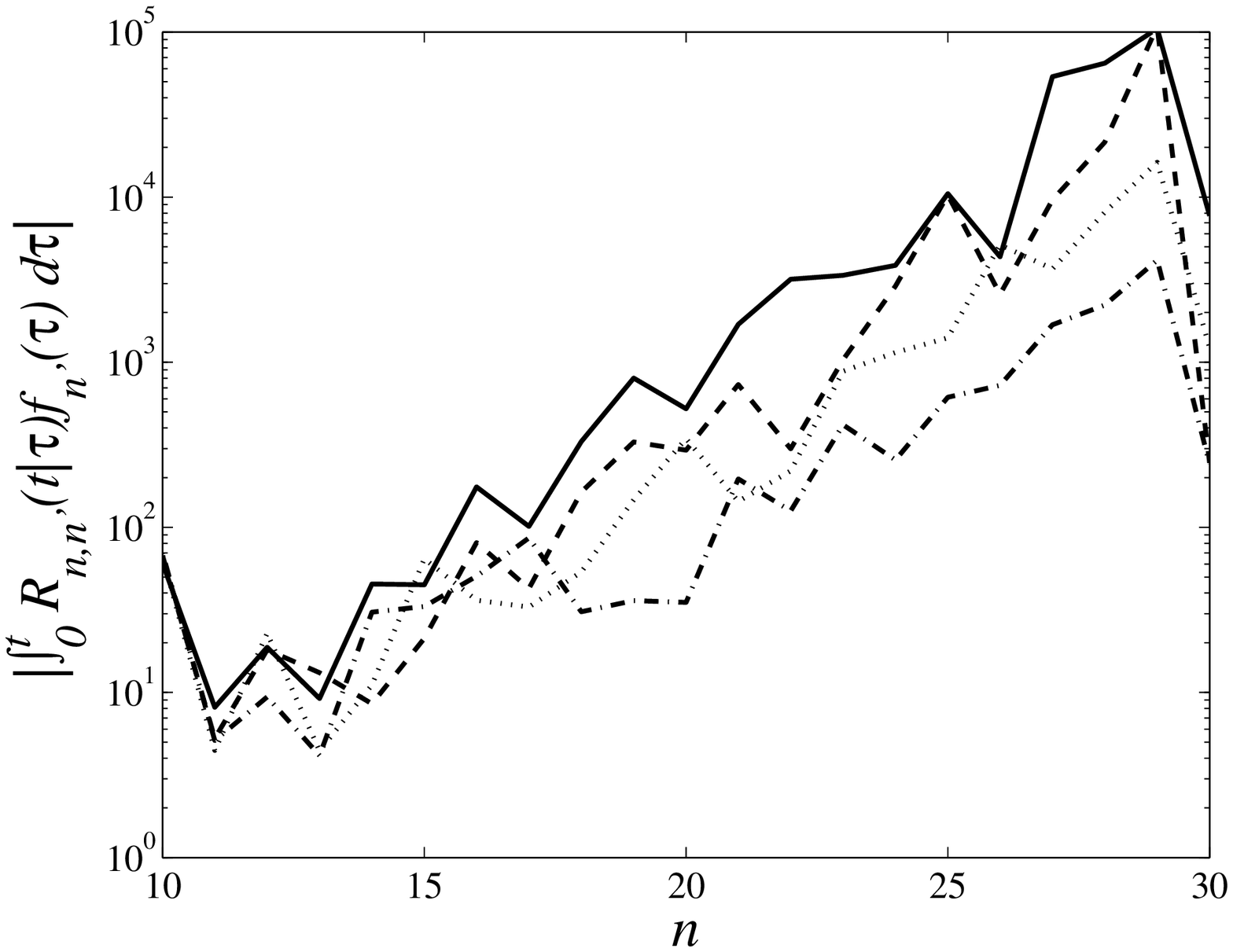}
\caption{Panel a: The fate of the modulus of the initial value term, averaged over
$2,000$  realizations. Shown
are shells 15, 17, 19, 21, 23 and 25 (top to bottom from the left-most
side). 
At time $t=0$ their
relative amplitudes agree with the scaling exponent of the
{\em active} field. As time progresses the decaying field switches
to the relative amplitudes which agree with the scaling exponent
of the {\em passive} field. Panel b: The modulus of four realizations of the integral
as a function of $n$, for time $t=3\times 10^{-4}$ (in unit of
$\tau_{10}$). Note that both terms of the
RHS of Eq. (\ref{boft}) exhibit the same {\em leading} scaling
behavior. This is canceled exactly as is demonstrated in Fig. \ref{cancel}}
\label{twoterms}
\end{figure}

\begin{figure}
\centering
\includegraphics[width=.45\textwidth]{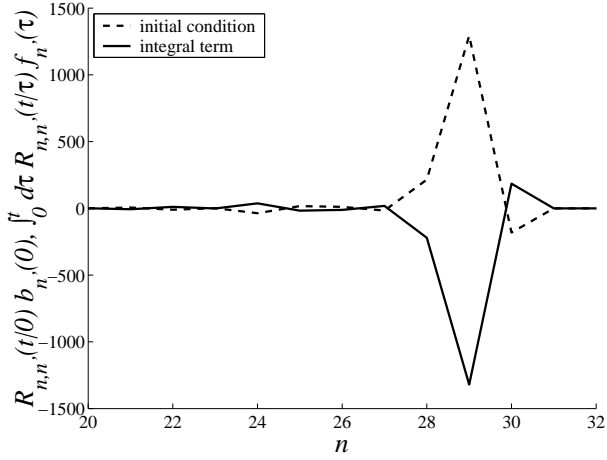}
\caption{the real part of the initial value term (dashed line) and
integral term (solid line) as a function of $k_n$. This is a demonstration
of the cancellation of the leading order term in favor of the subleading
one. The imaginary parts behave in the same way.}
\label{cancel}
\end{figure}

To gain insight into this interesting situation we note that the
analog of Eq.~(\ref{phioft}) describes the dynamics of our active
field $b_n$:
\begin{equation}
b_n(t) = R_{n,n'}(t|0)b_{n'}(t=0)
+\int_0^t d\tau R_{n,n_f}(t|\tau)f_{n_f}(\tau)\ , \label{boft}
\end{equation}
with an obvious re-definition of the present operator $R_{n,n'}$.
It is very revealing to examine the time dependence of the two
terms on the RHS of this equation. We measure time in units
of the eddy turn over time of the forcing shell 10. This is
defined as $\tau_{10}\equiv 
[k_{10} \sqrt {\langle |u_{10}|^2 \rangle}]^{-1}\approx 3.35 $.
We will examine a forced system
which began running at $t=-\infty$,  denoting a generic
time as $t=0$. In Fig. \ref{twoterms} panel a
we show the time-dependence of the first term for 6 values of $n$
in the inertial interval. We see that the initial conditions
represent, as expected, a 'standard' spectrum in which the
amplitude $b_n$ decreases as a function of $n$. As time proceeds,
the decaying term cannot recognize its being `active' from being
`passive', and it switches rapidly to the Statistically Preserved
Structure of the decaying passive field, characterized by a negative
exponent. If it were not for the second term on the RHS of
Eq. (\ref{boft}), then $b_n$ would have landed on the same
solution as $q_n$. What about the second term? In panel b of
Fig. \ref{twoterms} we show the $n$ dependence of the
term at time $t=3\times 10^{-4}$. We see that also this term agrees,
in its $n$ dependence, with the negative exponent of the
passive field. Yet, the LHS $b_n(t)$ fluctuates around
{\em decreasing} amplitudes as $n$ increases, meaning that
the leading (negative) exponent exactly cancels between
the two terms on the RHS of Eq. (\ref{boft}). We demonstrate this
cancellation in Fig. \ref{cancel}. There we plot the real
parts of the initial value term and the integral term
at time $t= 0.3$. We see that the two terms cancel each other.
The imaginary parts exhibit the same behavior.

Next we need
to understand this cancellation from the analysis of the
equations of motion. With this analysis we will also show
that the solution on which $b_n(t)$ is landing is also
a Statistically Preserved Structure of the decaying passive
field, albeit with a sub-leading scaling exponent.
\subsection{Statistically Preserved Structures of the passive
field}
In the next subsection we will show that the velocity
field attains a scaling solution with $\zeta_3=1$:
\begin{equation}
S_3(k_n)\equiv 
\Im<u_{n-1}u_{n}u_{n+1}^*>_f \sim k_n^{-1} \ .
\label{S3ofk}
\end{equation}
where $\Im$ denotes the imaginary
part. In this section we will assume this, and examine what are the
scaling solutions that agree with the existence of a 2nd order
Statistically Preserved Structure for the passive field.
We are not going to compute the anomalous scaling exponent exactly,
but rather obtain their dimensional estimates. Since we are after a
radically different apparent behavior, small numerical corrections
are not our main concern. With this caveat in mind,
we can calculate the exponent $\beta_3$ characterizing the third order
structure function 
\begin{equation}
P_3(k_n)\equiv \Im \langle q_{n-1}q_n q^*_{n+1}\rangle \sim k_n^{-\beta_3} \ .
\label{P3ofk}
\end{equation}
The condition for the existence of the 2nd order Statistically Preserved
Structure is
\begin{equation}
\frac{d}{dt}\langle |q_n|^2\rangle =0\ , \quad \forall n \ ,
\label{condzero}
\end{equation}
in the inviscid limit. Using Eq.~(\ref{sabraq})
this condition generates a number of third order quantities that need
to be analyzed first. Denote therefore
\begin{eqnarray}
Q_{3,1}(k_n) &\equiv& \Im<u_{n-1}q_{n}q_{n+1}^*>\ , \nonumber\\
Q_{3,2}(k_n) &\equiv& \Im<q_{n-1}u_{n}q_{n+1}^*>\ , \nonumber\\
Q_{3,3}(k_n) &\equiv& \Im<q_{n-1}q_{n}u_{n+1}^*>\  . \label{Qdef}
\end{eqnarray}
In order to construct scaling solutions for these objects, Dimensional
consideration imply that the
fields involved in the averages above have scalings $u_n\propto
\lambda^{- n/3}$ and $q_n \propto \lambda^{-\beta_3 n/3}$. We infer
the expressions
\begin{eqnarray}
Q_{3,1}(k_n)&=&|q_0|^2|u_0|k_n^{-(2\beta_3+1)/3}
\lambda^{-(\beta_3-1)/3} \ ,\nonumber\\
Q_{3,2}(k_n)&=&|q_0|^2|u_0|k_n^{-(2\beta_3+1)/3}\ , \nonumber\\
Q_{3,3}(k_n)&=&|q_0|^2|u_0|k_n^{-(2\beta_3+1)n/3}
\lambda^{(\beta_3-1)/3} \ .
\end{eqnarray}
We can thus rewrite
\begin{eqnarray}
Q_{3,1}(k_n) &\equiv& \lambda^{-(\beta_3-1)/3} \tilde Q_3(k_n)\
,\label{redefQ31}\\
Q_{3,2}(k_n) &\equiv& \tilde Q_3(k_n)\ ,\label{redefQ32}\\
Q_{3,3}(k_n) &\equiv& \lambda^{(\beta_3-1)/3} \tilde Q_3(k_n)\
,\label{redefQ33}
\end{eqnarray}
where $\tilde Q_3(k_n)$ scales like
\begin{equation}
\tilde Q_3(k_n)\equiv \lambda^{-(2\beta_3+1)n/3}\ .
\label{scalingQ3}
\end{equation}
having these definitions in mind we derive, by demanding
Eq.~(\ref{condzero})
and substituting the scaling form of $\tilde Q_3$ Eq. (\ref{scalingQ3}),
the equation
\begin{widetext}
\begin{equation}
\tilde{a}\lambda^{-(2\beta_3-2)/3}
(1 - \lambda^{(\beta_3-1)/3})+
\tilde{b}(\lambda^{-(\beta_3-1)/3}-\lambda^{(\beta_3-1)/3})
+\tilde{c}\lambda^{(2\beta_3-2)/3}
(\lambda^{-(\beta_3-1)/3}-1) = 0\ .
\label{eqeta2}
\end{equation}
\end{widetext}
This is  a fourth order polynomial in $\lambda^{(\beta_3-1)/3}$. The
four roots are
\begin{equation}
\lambda^{(\beta_3-1)/3} = \left\{
\begin{array}{l}
1\ ,\\
\lambda^{-\alpha}\ ,\\
\pm \lambda^{-\alpha/2}\ .
\end{array}\right.
\label{passscal2d}
\end{equation}
Here three of the roots correspond to a priori physical solutions:
\begin{equation}
\beta_3 = \left\{
\begin{array}{l}
1\ ,\\
1 - 3\alpha/2\ ,\\
1 - 3\alpha\ .
\end{array}\right.
\label{solbeta}
\end{equation}
In our simulations with $\alpha=2$ these results are $\beta_3=1$, -2 and -5
respectively. This is in agreement with spectral exponents $\beta_2$
of the order of (neglecting anomalies) $\beta_2=2/3, -4/3, -10/3$.
To know which of these is physical, we need to consider the fluxes
supported by these solutions.

The only flux that is relevant for the passive field is the magnetic
helicity. For the case considered here with $\delta>1$ 
it can be conveniently computed at the shell
$M$ by evaluating
\begin{equation}
\Phi^H_M \equiv - \frac{1}{2}\frac{d}{dt}\sum_{n=0}^M
\left<\frac{|q_n|^2}{k_n^\alpha}\right>
\ .\label{hflux}
\end{equation}
Using the equations of motion to evaluate this object we find
\begin{widetext}
\begin{eqnarray}
\Phi^H_M = &-& \delta_m
\Big[ k_{M+1}^{1-\alpha}
\big(\Im<q_{M}u_{M+1}q_{M+2}^*> - \Im<q_{M}q_{M+1}u_{M+2}^*>\big)\nonumber\\
&+& k_M^{1-\alpha}\big(\Im<q_{M-1}u_{M}q_{M+1}^*>
+ \Im<u_{M-1}q_{M}q_{M+1}^*>\big)\Big]\ . \label{exphflux}
\end{eqnarray}
\end{widetext}
We can evaluate now the magnetic helicity flux for the three
scaling solutions (\ref{solbeta}). We find
\begin{equation}
\Phi^H_M \propto \left\{
\begin{array}{l}
\lambda^{-\alpha M}\ ,\\
1\ ,\\
\lambda^{\alpha M} \ .
\end{array}\right.
\label{solpassiveflux}
\end{equation}
We conclude that the third solution is unphysical, since it supports
a flux that diverges with $M$. The first two solutions are allowed.
With $\beta_3=-2$ we get a constant flux; this is the leading
scaling solution, and is indeed realized in the simulations. The
solution $\beta_3=1$ is subleading, it is associated with a decaying
flux, and is asymptotically allowed. It is not observed in the passive
field simulations simply because it is subleading.

Our main point will be that the {\em active} field will in its turn
land on the subleading Statistically Preserved Structure
because the additional conservation laws exclude the leading
one. We demonstrate this phenomenon in the next subsection.
\subsection{Why does the active field fall on a subleading zero mode?}

If we accept the general philosophy that active fields exhibit
scaling behaviors that are determined by the zero modes of the
auxiliary passive fields, then we should explain here why
in the present case the active field avoids the {\em leading}
zero mode, and appears to land on the sub-leading one. The answer
is hidden of course in the conservation laws, as we expose now.

We first repeat the analysis performed in the previous
section to find the consequence of the equation
\begin{equation}
\frac{d}{dt} <|b_n|^2> = 0\ ,
\label{ddtb2}
\end{equation}
or, equivalently,
\begin{equation}
\frac{d}{dt} <|u_n|^2> = 0\ .
\label{ddtu2}
\end{equation}
Using now the definitions
\begin{eqnarray}
B_3(k_n) &=& \Im \langle b_{n-1} b_n b^*_{n+1}\rangle \sim
k_n^{-\eta_3}\nonumber\\
Q_3(k_n)&\equiv&  k_n^{-(2\eta_3+\zeta_3)/3} \ , \
\end{eqnarray}
we obtain an equation that is analogous to Eq.~(\ref{eqeta2}),
\begin{widetext}
\begin{equation}
\tilde{a}\lambda^{1-(2\eta_3+\zeta_3)/3}
(1 - \lambda^{(\eta_3-\zeta_3)/3})+
\tilde{b}(\lambda^{-(\eta_3-\zeta_3)/3}-\lambda^{(\eta_3-\zeta_3)/3})
+\tilde{c}\lambda^{-1+(2\eta_3+\zeta_3)/3}
(\lambda^{-(\eta_3-\zeta_3)/3}-1) = 0\ .
\label{eqeta}
\end{equation}
\end{widetext}
This is a fourth degree polynomial for $\lambda^{\eta_3/3}$ if $\zeta_3$
is known. Obviously,
if we simply substituted here $\zeta_3=1$ we would get the same predictions
for $\eta_3$ as obtained for $\beta_3$ in Eq. (\ref{solbeta}). However, we
have
in this case an important additional constraint that is absent in the case
of the passive field, which can be inferred from the additional
conservation equation
\begin{equation}
\frac{d}{dt} \Re<u_n^*b_n> = 0
\label{ddtub}
\end{equation}
Repeating the analysis as above, and introducing a  new object
$A_3(k_n)\equiv k_n^{-(\eta_3+2\zeta_3)}$ yields the two equations
\begin{eqnarray}
&&a\lambda B_3(k_{n+1}) + b B_3(k_n) + c\lambda^{-1}B_3(k_{n-1}) =0\ ,
\label{eqP3}\\
&&a\lambda^{1-(\eta_3-\zeta_3)/3} A_3(k_{n+1})+ b A_3(k_n)\nonumber\\
&&\quad + c\lambda^{-1+(\eta_3-\zeta_3)/3}A_3(k_{n-1})=0\ ,
\label{eqR3}
\end{eqnarray}
Solving this system together with Eq. (\ref{eqeta}) yields the scaling
exponents
\begin{equation}
\zeta_3,\:\eta_3 = \left\{\begin{array}{l@{\quad}l}1&\\
1+ \log_{\lambda}(a/c)& \ .
\end{array}
\right.
\label{scalingsol}
\end{equation}
It is easy to check that, among the four possible combinations of
this equation, the only solution allowed by Eq. (\ref{eqeta}) is
\begin{equation}
\zeta_3=\eta_3=1\ .
\label{zetaeta}
\end{equation}
We thus conclude that as far as the active field
is considered, the additional conservation law rules out the leading
zero mode of the passive problem, leaving us only with the subleading mode
which is observed in the simulations.
\section{Summary and Conclusions}
\label{summary}
In this paper we considered the correspondence between the statistics
of active fields that are advected by a turbulent velocity field,
and the statistics of an auxiliary passive field that is advected
by the same velocity, but does not affect it. The two
examples were akin to turbulent convection and to magneto-hydrodynamics
respectively. In the first example the conserved variables for
the equations of the passive and active fields are the same. For the
second example the active problem exhibits additional conservation laws.
This was shown to be very significant in determining the respective
statistical physics of the two problems.

The two examples appear very different in superficial
examination. In the first example the even-order statistics of the
passive and active fields turned out to be the same up to a single
multiplicative factor $\beta$, common to all orders. The forced structure
functions of the active field scale with exactly the same exponents
as the passive field, which in turn are dominated by the leading zero modes
of the decaying problem. We analyzed in detail the source of the
multiplicative factor $\beta$ and showed that it stems from the
additional correlation effects between the forcing and the velocity field
that are absent in the passive case.
Nevertheless these correlation effects do not cause a change in the
scaling exponents. The general lesson that we would propose on the
basis of this example is that whenever there exist a problem in
which the equation of motion of the active field does not satisfy
additional conservation
laws compared to the passive case, the former field
will exhibit structure functions that are dominated by the leading
zero modes of the latter.  This point is also pertinent to the second
example.
Here the active equations possess additional conservation laws,
and indeed the active and passive fields exhibit different scaling
exponents.
Nevertheless we argued that the structure functions of the active field
are still dominated by the zero modes of the passive problem, but
not the leading ones. The additional conservation laws results in
exact cancellations in the contributions of the leading zero modes,
and the active problem lands on the next allowed sub-leading zero mode
of the passive problem.

As a generalization, consider then a sufficiently
turbulent velocity field which advects an active field, scalar or
vector, which in its turn is forced by a force having a compact
support in $\B.k$ space. An auxiliary passive field which is advected
by the same velocity field can be employed to find the
zero modes of the operator involved in the passive decay
problem. On the basis of the intuition gained
with the examples presented above, we offer the following tentative
conjecture: the forced structure function of the active field
will exhibit scaling exponents that are the scaling exponents of the
aforementioned 
zero modes. Whenever the conservation laws of the active and passive
problems coincide, these will be the exponents of the leading zero modes.
When the active problem has additional conservation laws, these will be
the next-leading zero modes, as allowed by the conservation laws.

Finally, we need to consider the relation of our shell models
to the physical problems and the PDE's that motivate these models. It is
important to test the
conjecture stated here in that context. In light of the above discussion
we expect that much of what has been found here will translate literally
to the continuous problems. After all, the crucial aspects are the linearity
of the advection equation, and the existence of conservation laws. These
are unchanged in the continuous problems. Of course, one can expect
many more numerical difficulties, especially due to the role of angles
in the multi-point correlation functions. Nevertheless, the idea that
the understanding of the anomalous scaling exponent of active fields boils
down to the analysis of eigenfunctions of a linear operator is expected
to hold verbatim.
\acknowledgments
This work has been supported in part by the
Research Grants Council of Hong Kong SAR, China
(CUHK 4119/98P and CUHK 4286/00P), the European Commission
under a TMR grant, by the Minerva Foundation, Munich, Germany, the German
Israeli Foundation, and the Naftali and Anna Backenroth-Bronicki Fund for
Research in Chaos and Complexity. TG thanks the Israeli Council for Higher
Education and the Feinberg 
postdoctoral Fellowships program at the WIS for financial support.

\end{document}